# Sparse electrophysiological source imaging predicts aging-related gait speed slowing.

**Vega-Hernández, Mayrim[1,2], Galán-García, Lídice[2], Pérez-Hidalgo-Gato, Jhoanna[2], Ontivero-Ortega, Marlis[2], García-Agustin, Daysi[3], García-Reyes, Ronaldo[2], Bosch-Bayard, Jorge[4], Marinazzo, Daniele[5], Martínez-Montes, Eduardo[2,*] and Valdés-Sosa, Pedro A.[1,2,*]**

[1]The Clinical Hospital of Chengdu Brain Science Institute, MOE Key Lab for Neuroinformation, University of Electronic Science and Technology of China, Chengdu, China.
[2]Cuban Center for Neurosciences, Havana, Cuba.
[3]Cuban Centre for Longevity, Ageing and Health Studies, Cuba
[4]Montreal Neurological Institute, Canada.;
[5]Faculty of Psychology and Educational Sciences, Department of Data Analysis, Ghent University.

* Corresponding author at:

Cuban Center for Neurosciences, Human Brain Mapping Division, Havana, Cuba.

The Clinical Hospital of Chengdu Brain Science Institute, MOE Key Lab for Neuroinformation, University of Electronic Science and Technology of China, Chengdu, China.

E-mail address: eduardo@cneuro.edu.cu; pedro.valdes@neuroinformatics-collaboratory.org.

## HIGHLIGHTS

- Electrophysiological Source Imaging (ESI) predicts the individual gait speed slowing measured with 4-Meter Walk Test in elders.
- Best predictors of gait slowing with aging are a combination of orbitofrontal and temporal Theta activation and connectivity.
- Our novel sparse non-negative smooth ESI achieved better gait speed prediction than state-of-the-art methods.

## ABSTRACT

**Objective:** We seek stable Electrophysiological Source Imaging (ESI) biomarkers associated with Gait Speed (GS) as a measure of functional decline. Towards this end we determine the predictive value of ESI activation and connectivity patterns of resting-state EEG Theta rhythm on physical performance decline measured by a slowing GS in aging individuals.
**Methods:** As potential biomarkers related to GS changes, we estimate ESI using flexible sparse/smooth/non-negative models (NN-SLASSO), from which activation ESI (aESI) and connectivity ESI (cESI) features are selected using the Stable Sparse Classifier method.
**Results and Conclusions:** Novel sparse aESI models outperformed traditional methods such as the LORETA family. The models combining aESI and cESI features improved the predictability of GS changes. Selected biomarkers from activation/connectivity patterns were localized to orbitofrontal and temporal cortical regions.
**Significance:** The proposed methodology contributes to understanding the activation and connectivity of ESI complex patterns related to GS, providing potential biomarker features for GS slowing. Given the known relationship between GS decline and cognitive impairment, this preliminary work suggests it might be applied to other, more complex measures of healthy and pathological aging. Importantly, it might allow an ESI-based evaluation of rehabilitation programs.

**Abbreviations:** ESI, Electrophysiological Source Imaging; GS, Gait Speed; aESI, activation ESI; cESI, connectivity ESI; AMNR, Active-set Modified Newton Raphson; MPLS, Multiple Penalized Least Squares; DGS, Delta Gait Speed or gait speed variability; acESI, activation/connectivity ESI; ICC, intraclass correlation coefficient; ROIs, regions of interest; SSC, Stable Sparse Classifiers.

## 1. Introduction

The increase in life expectancy and the decline in the birth rate in recent decades have led to an aging population worldwide with an increase in age-related diseases and disabilities. A key feature of aging is functional decline, which has far-reaching consequences for individuals, families, caregivers, and society. According to the WHO, around 55 million people worldwide live with dementia, with over 60% in low- and middle-income countries. This number is expected to rise to 78 million in 2030 and 139 million by 2050. In 2019, informal caregivers (mostly family and friends) spent around 5 hours, on average daily, to care for people with dementia. This produces significant physical, emotional, and financial stress for the families and caregivers, with a substantial economic impact (https://www.who.int/news-room/fact-sheets/detail/dementia). Although chronological age is strongly associated with age-related functional decline, there are significant differences between individuals of the same age. Informative and reliable biomarkers that predict individuals at risk of developing functional loss are needed for personalized medicine strategies to promote healthy aging. To discover these biomarkers, an ideal alternative is to use longitudinal studies to estimate the evolution of these deficits individually.

Recent studies suggest that the EEG can be a cost-effective tool for assessing functional decline and cognitive changes associated with aging and other age-related pathologies. A relationship between quantitative EEG (e.g., power spectrum) changes across different frequency bands and cognitive impairment has been seminally reported by Prichep and colleagues (Prichep et al., 2005). Musaeusa et al. (2018) found that increased Theta power on the EEG is an early marker of cognitive decline in dementia due to Alzheimer's disease. Quantitative EEG and functional EEG connectivity as biomarkers of cognitive impairment is still an active area of research (Smailovic et al., 2018; Smailovic & Jelic, 2019). Although less explored, estimation of the EEG source activation and connectivity using Electrophysiological Source Imaging (ESI) is known to provide relevant information about the functional organization of the brain and interregional communication (Paz-Linares et al., 2023; He et al., 2019; Van de Steen et al., 2019). ESI has also been used to investigate changes in brain function associated with both healthy aging and neurodegenerative diseases, including Alzheimer's disease and Parkinson's disease (Jabès et al., 2021; Meghdadi et al., 2021; Muthuraman et al., 2014). Babiloni et al. (2018) investigated the functional connectivity of alpha rhythm sources in patients with mild cognitive impairment (MCI), while Vecchio and colleagues used EEG connectivity and graph theory combined with Apolipoprotein E to evaluate the risk of MCI progression (Vecchio et al., 2018). Hata and colleagues established that cerebrospinal fluid biomarkers of Alzheimer's Disease correlate with EEG source parameters (mean activation and lagged phase synchronization) assessed by exact low-resolution electromagnetic tomography (eLORETA) (Hata et al., 2017).

In this work, we also pursue establishing EEG biomarkers of functional decline in aging in the framework of a longitudinal study, as guided by three considerations.

1- Tackling EEG biomarker discovery for complex cognitive behavior measured by neuropsychological and cognitive tests involves high costs and specialized personnel. We instead focus on indirect measurements that are easier to obtain, specifically Gait Speed (GS). GS has been widely used in clinical practice as a predictor of adverse outcomes (Kahya et al., 2019; Afilalo et al., 2016) and an important biomarker of human health (Nascimento et al., 2022; Skillbäck et al., 2021). This is a measurement that can be performed by non-professional trained staff and is one of the most attractive physical biomarkers of cognitive decline (Nascimento et al., 2022; Cohen & Verghese, 2019; Kahya et al., 2019). Some studies using GS and cognitive tasks have found consistent poor performance in various independent gait domains preceding cognitive decline and incident dementia (e.g., Skillbäck et al., 2021; Darweesh et al., 2019; Dumurgier et al., 2017). GS is an early proxy of aging-related cognitive decline (Nascimento et al., 2022; Collyer et al., 2022; Skillbäck et al., 2021; Zhou et al., 2021; Öhlin et al., 2020, 2021; Montero-odasso et al., 2020). Some works have reported the study of the relation between GS and EEG (De Sanctis et al., 2023; Min et al., 2022; Nordin et al., 2020), generally using the EEG spectrum. Previous longitudinal studies on GS and cognition have indicated that GS changes occurred as early as 7 to 12 years before dementia development (Dumurgier et al., 2017). Skillbäck et al. (2021) demonstrated that accelerated GS decline preceded cognitive decline and indicated faster GS decline when conversion into detectable cognitive deficiency is imminent.

2- Though there are many studies on the EEG sources of cognitive abilities (Babiloni et al., 2018; Vecchio et al., 2018; Hata et al., 2017), which highly influences the measures of brain connectivity derived from them, it is also noticeable that, to our knowledge, none of these current studies refer to GS.

3- Furthermore, to date, biomarker selection has been hampered by the lack of robustness of the selected model against small perturbations of the data set. We leverage the use of resampling techniques to evaluate the stability of models when performing penalized regressions, which opens the way to obtaining a stable estimation of the biomarkers (Bosch-Bayard et al., 2018).

We address these three aspects with a method that combines two main steps:

- The estimation of potential biomarkers from the EEG based on the sparse activation and connectivity ESI.
- The identification of those biomarkers from these ESI-derived features that are stable and accurate predictors of GS.

Specifically, we used data from a longitudinal study in healthy older adults, in which the walking or gait speed (GS) was measured in two different sessions separated by an interval of 6 years and the resting-state EEG was acquired only in the last assessment. However, the methodology introduced here is quite general and can be applied based on any ESI method to predict any functional change.

## 2. Materials and methods

### 2.1 General methodology

The method used to predict GS from activation and/or connectivity ESI is shown in Figure 1. In short, Step 0 details data acquisition and pre-processing. Step 1 represents the computation of specific inverse methods and ESI features, the definition of nodes for the connectivity analysis, and the standardization of the final features. Step 2 includes using classifiers for biomarker/variable selection from the features obtained from Step 1. The following subsections will explain the details of each block for the presented study.

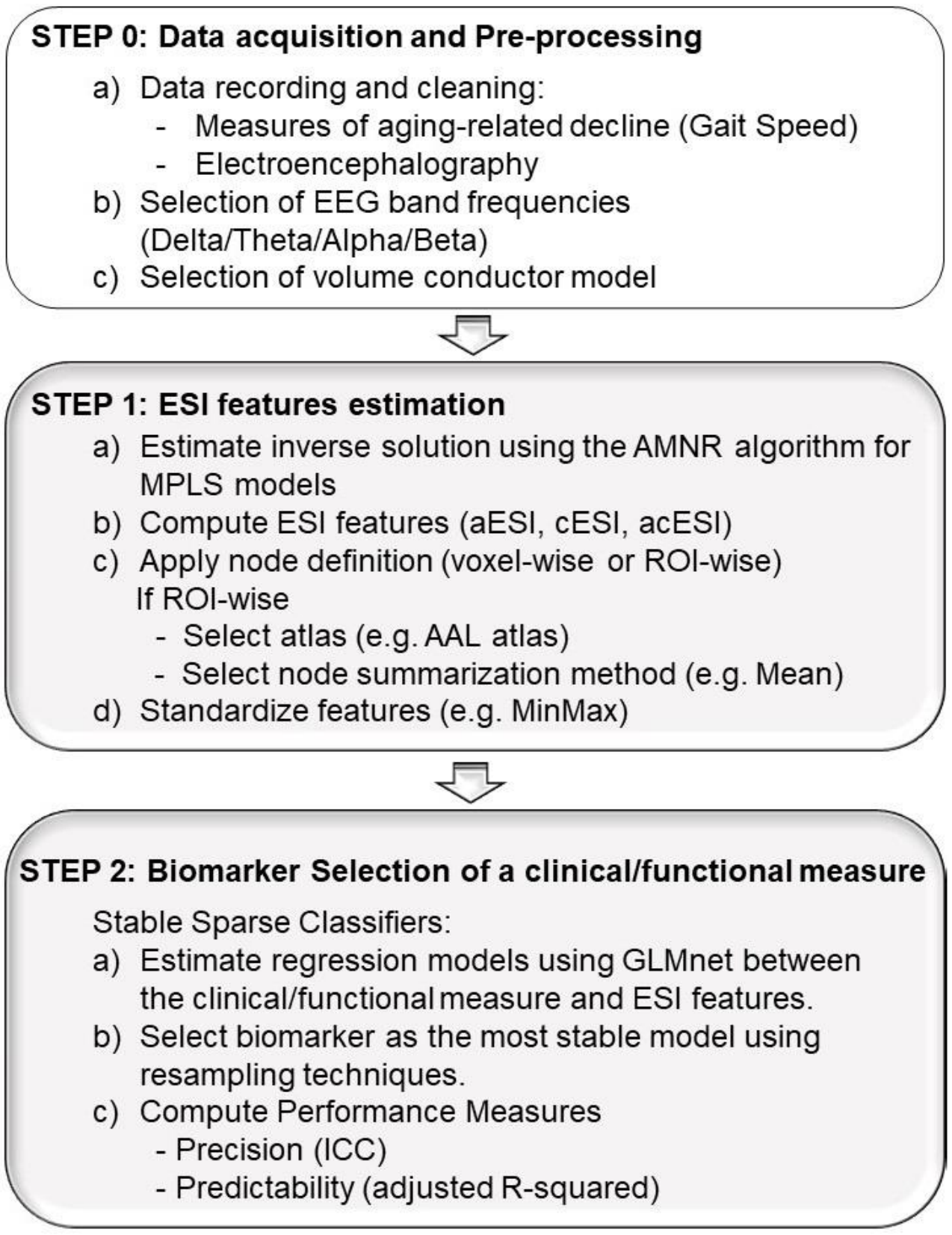

**Fig. 1. Schematic representation of the proposed methodology to use features derived from Electrophysiological Source Imaging as predictors of changes in a clinical/functional measure of cognitive decline (in our case changes in gait speed).**

### 2.2 STEP 0: Data acquisition and pre-processing

a) The dataset was a subsample from a longitudinal study on healthy aging, conducted in Havana, Cuba from 2007 to 2016 as part of a project registered in the Centro de Investigaciones sobre Longevidad, Envejecimiento y Salud (CITED) named “Bases electrofisiológicas de las alteraciones en la movilidad de los adultos mayores” (Code: 17231705). For our study, we selected a subsample of 66 community-dwelling adults who regularly practiced mild-to-moderate exercise in their communities. Gait speed (GS) was evaluated in 2010 and 2016, while EEG was only recorded at the last time point(García-Agustin et al., 2020). The age range of the participants varies from 64 to 94 years old (mean age: 78.73 ± 6.69 years).
Details of the measurements are as follows:
**Gait Speed recording:** The 4-Meter Walk Test (4MWT) quantified the GS with high test-retest reliability and validity (Peters et al., 2013). This technique requires a stopwatch and a facility hallway and can be performed by non-professional trained staff (Tzemah-Shahar et al., 2022; Fritz & Lusardi, 2009). The measurement consisted of quantifying the time spent to cover 4 meters at a normal pace and was expressed in meters per second (m/s). Individual gait speed variability (Delta Gait Speed, DGS) was calculated for each subject as the difference between the two measured gait speeds (in 2010 and 2016) relative to the initial GS.

**EEG recordings:** The EEG data were recorded using the Neuronic EEG recording system MEDICID 5 (Neuronic S.A.). Nineteen Ag/AgCl electrodes were placed on the scalp according to the international 10/20 electrode placement system, taking the linked earlobes as the reference and keeping electrode impedance below 10 kOhm. The following acquisition parameters were employed: gain of 10,000; bandpass filtering between 0.3-30 Hz, notch filtering at 60 Hz, sampling interval of 5 msec and environmental temperature of approximately 23ºC. The EEG was acquired at rest for 10 minutes while the participants were asked to close and open their eyes at different times to explore reactivity and avoid drowsiness. **EEG cleaning:** Epoch selection during the eyes-closed resting state was performed offline using Neuronic EEG analysis (Neuronic SA), obtaining 18 to 24 artifact-free segments of 2.56 s for each subject by visual inspection of two experienced neurophysiologists.

b) The Fast Fourier Transform (FFT) was applied to those artifact-free EEG epochs to find the complex EEG amplitude spectra in the frequency domain. These values were the input to calculate the Electrophysiological Source Imaging (ESI) for all 49 frequencies from 0.39 to 19.14 Hz (step of 0.39 Hz). The absolute values of each ESI were averaged within the Theta band, defined from 5.47 to 7.03 Hz (Babiloni et al., 2020; Bosch-Bayard et al., 2001).

c) A standard template head model was used for estimating ESI on all subjects. The generating sources were defined over a cortical surface mesh containing 5656 vertices on brain regions that are physiologically feasible EEG generators (e.g., avoiding vertices in corpus callosum). Using a three-sphere piece-wise homogenous and isotropic head model, the projection matrix (i.e. the Electric Lead Field) was computed for the standard positions of the array of 19 electrodes from the 10/20 system in Neuronic Source Localizer (Riera & Fuentes, 1998).

## 2.3 STEP 1: Estimation of ESI features

a) ESI was estimated using the general theoretical framework of the Multiple Penalized Least Squares (MPLS) model:

$$\hat{\mathbf{J}} = \underset{\mathbf{J}}{\operatorname{argmin}}\{\|\mathbf{V} - \mathbf{KJ}\|_2^2 + \Psi(\mathbf{J})\} \quad \text{with} \quad \Psi(\mathbf{J}) = \sum_{r=1}^{R} \lambda_r \sum_{i=1}^{N_r} \boldsymbol{g}^{(r)}\left(\left|\theta_i^{(r)}\right|\right)$$

where $\mathbf{V}$ represents the EEG data matrix, $\mathbf{J}$ is the unknown primary current density reflecting source activity, $\mathbf{K}$ is the Electric Lead Field, $\lambda_r$ represent regularization parameters and the penalty functions $\boldsymbol{g}^{(r)}: \mathbb{R} \mapsto \mathbb{R}$, are symmetric, non-negative, non-decreasing and continuous over $(0, +\infty)$, with the variables $\boldsymbol{\theta}^{(r)}$ being a linear combination of the unknown parameters $\mathbf{J}$. This model allows a common formulation of a large variety of solutions. In this paper we used three solutions: i) a version of Ridge theoretically equivalent to a Low-Resolution Electromagnetic Tomography (LORETA) (Pascual-Marqui et al., 1994), which includes only an $\ell$2-norm penalty to enforce a smooth solution; ii) the more flexible Elastic Net L (ENET L) model that allows finding solutions with different degrees of sparsity and smoothness (Vega-Hernández et al., 2008, 2019; Haufe et al., 2008, 2011; Sohrabpour et al., 2016); iii) the Non-Negative Smooth LASSO (NN-LASSO), which imposes non-negativity constraints to smooth/sparse solutions. The mathematical formulation of the penalty functions of each method is provided in Table 1. In this context, using efficient algorithms to estimate these flexible MPLS models is critical to address real-world EEG data´s complexities effectively. Here we used the Active-Set Modified Newton-Raphson (AMNR) algorithm that allows efficient estimation of solutions combining smoothness and sparseness but also including sign constraints in a natural way (Vega-Hernández et al., 2019) (Docker with Matlab code available at https://portal.cbrain.mcgill.ca/).

**Table 1.** Penalty terms for the methods LORETA, Elastic Net (ENET L) and Nonnegative Smooth LASSO (NN-SLASSO). $\mathbf{L}$ is the matrix of second differences, or any other roughness operator, and $\lambda_r$ represent the regularization parameters.

| Name | Penalty term $\boldsymbol{\Psi}(\boldsymbol{\beta})$ | Function definitions for all models |
|---|---|---|
| LORETA | $\lambda_1 \sum_{i=1}^{\mathbf{p}} \boldsymbol{g}^{(1)}\left(\left|\boldsymbol{\theta}_i^{(1)}\right|\right)$ | $\boldsymbol{g}^{(1)}(\theta) = \theta^2$; $\boldsymbol{\theta}^{(1)} = \mathbf{LJ}$ |
| ENET L | $\lambda_1 \sum_{i=1}^{\mathbf{p}} \boldsymbol{g}^{(1)}\left(\left|\boldsymbol{\theta}_i^{(1)}\right|\right) + \lambda_2 \sum_{i=1}^{p} \boldsymbol{g}^{(2)}\left(\left|\boldsymbol{\theta}_i^{(2)}\right|\right)$ | $\boldsymbol{g}^{(2)}(\theta) = |\theta|$; $\boldsymbol{\theta}^{(2)} = \mathbf{LJ}$ |
| NN-SLASSO | $\lambda_1 \sum_{i=1}^{\mathbf{p}} \boldsymbol{g}^{(1)}\left(\left|\boldsymbol{\theta}_i^{(1)}\right|\right) + \lambda_2 \sum_{i=1}^{p} \boldsymbol{g}^{(3)}\left(\left|\boldsymbol{\theta}_i^{(3)}\right|\right)$ | $\boldsymbol{g}^{(3)}(\theta) = |\theta|$; $\boldsymbol{\theta}^{(3)} = \mathbf{J}$ ***subject to*** $\mathbf{J} \geq \mathbf{0}$ |

b) The absolute value of the estimated (frequency-domain) complex solution ($\mathbf{J}$) from each method will be called the activation ESI (aESI), which is the first feature to be assessed as a potential biomarker. A second ESI-derived feature would reflect brain connectivity at the source level. In this study, we used the absolute value of the source cross-spectra, estimated as the covariances of complex solutions in the Theta band along segments of eyes-closed resting-state EEG. The symmetric covariance matrix contains information on the activation strength (in the variances, in the diagonal elements) and on the connectivity magnitude (in the off-diagonal elements). We used a vector of absolute values of all pairwise covariances as the connectivity ESI (cESI) feature. A third feature allows exploring the combination of both information by forming a vector concatenating variances and the vector of cESI, which will be called activation/connectivity ESI (acESI).

c) In practice, the aESI (cESI) feature gives one value for each source generator (pair of source generators), i.e., they provide vertex-wise (or, in general, voxel-wise) magnitudes. In principle, the voxel-wise activations and connectivity have a high spatial resolution. However, it is known that they will present a higher risk of reflecting spurious activations and short-range connectivity due to partial volume/smoothing effects, making it difficult to assess their validity. In addition, processing all individual activations and pairwise connections is computationally demanding. Therefore, we here opted to explore the voxel-wise feature only for aESI. For all three ESI features (aESI, cESI, and acESI), we also reduced the dimensionality by summarizing them using regions of interest (ROIs) according to a specified anatomical atlas. This step might facilitate the interpretation of the selected biomarkers in terms of brain regions and not just individual voxels. In particular, we used the Automated Anatomical Labeling (AAL) segmentation from the Montreal Neurological Institute atlas (Tzourio-Mazoyer et al., 2002) for computing the activation and connectivity in 76 anatomical compartments as the mean value across all sources (vertices of the cortical mesh) included in each ROI (see labels in Fig. S1 and Table S1 of the Supplemental Material).

d) ESI-derived features will also reflect the large inter-subject variability typical of the EEG and ESI. Although there is no easy way to know if normalization of the individual data is necessary for finding group effects, we decided to use a Mix-Max standardization of all features, both voxel-wise or ROI-wise, to get normalized values for each subject in the range [0,1] and facilitate the implicit comparison when selecting the potential biomarker models.

## 2.4 STEP 2: Biomarker selection

A general methodology to select potential biomarker models from a pair of dependent and independent variables was presented by Bosch-Bayard and colleagues in the Stable Sparse Classifiers (SSC) procedure (Bosch-Bayard et al., 2018) (Docker with Matlab code available at https://portal.cbrain.mcgill.ca/) which is based on the following steps:

a) Use penalized regression as a variable selection method. In particular, the elastic-net model (Zou & Hastie, 2005) as implemented in the GLMnet package for Matlab (https://hastie.su.domains/glmnet_matlab/),which correspond to the optimization problem:

$$\hat{\boldsymbol{\beta}} = argmin\{\|\mathbf{Y} - \mathbf{X}\boldsymbol{\beta}\|_2^2 + (1-\gamma)\|\boldsymbol{\beta}\|_2^2 + \gamma\|\boldsymbol{\beta}\|_1\}$$

Here, $\mathbf{Y} \in \mathbb{R}^N$ is the dependent variable, in our case, the measure of aging-related decline (delta gait speed) for each of the $N$ subjects; $\mathbf{X}$ is the independent variable conforming a design matrix where each row represents the vector ESI-derived feature for each subject; $\boldsymbol{\beta} \in \mathbb{R}^p$ are the model parameters or regression coefficients for each of the $p$ elements of the ESI feature; $\gamma$ is the regularization parameter; and $\|\cdot\|_1$ and $\|\cdot\|_2$ represent the $\ell1$-norm and $\ell2$-norm, respectively.

b) Use quantitative indicators to assess the stability of the selected variables (significant regression coefficients). Subsampling the data in both dimensions -the observations and the variables- allows performing repeated regressions to estimate the variability of coefficients along each dimension. Variables selected on a significant fraction of the subsampled data after several iterations are considered significantly stable features. The principle is that if some features are consistently significant across all models, they can point to variables that are strong indicators of a potential stable biomarker. We reported the replicability of a variable as measured by the percentage of times that the variable is included in the selected models.

c) Evaluate the performance of the classifier using quality measures. The method calculates the receiver operator characteristic (ROC) and area under the curve for each model in each input subsampled data. This allows estimating the probabilistic distribution of the area under the ROC curve and therefore provides the possibility of combining the criteria of stability and precision for the selection of the model with statistically higher performance.
In addition, this methodology provided the adjusted R-square ($R^2$) and intraclass correlation coefficient (ICC) for the highest performing selected model. The first gives a general measure of goodness of fit and predictability. The ICC can be interpreted as a generalized measure of the precision of the model. The ICC estimator is computed as:

$$\text{ICC} = (MS_R - MS_W) / (MS_R + (k-1) MS_W)$$

where $MS_R$ represents the mean squares of the model, $MS_W$ the mean squares for residual sources of variance and $k$ is the number of measurements (Koo & Li, 2016). There are no standard values for acceptable reliability using ICC, but previous studies suggested that ICC values lower than 0.50 indicated poor reliability, values between 0.50 and 0.75 moderate reliability, values between 0.75 and 0.90 good reliability, and values higher than 0.90 excellent reliability (Koo & Li, 2016; Liljequist et al., 2019).

## 3. RESULTS

### 3.1 Data characterization

Table 2 details the statistical behavior of the longitudinal changes in the measured gait speed, i.e., the delta gait speed (DGS), as well as of the initial and final GS (measured in 2010 and 2016, respectively). The initial and final GS values are significantly different in the group of subjects (T-test for dependent samples: t (65) =10.36, p<<0.01).

**Table 2.** Descriptive statistics for gait speed (GS) variables. SD= standard deviation, CI = confidence interval (95%)

| Variable | Mean (±SD) | CI |
|---|---|---|
| **GS *initial*** | 1.036 (±0.28) | [0.968,1.103] |
| ***GS final*** | 0.753 (±0.21) | [0.701, 0.804] |
| **DGS** | -0.254 (±0.20) | [-0.303, -0.204] |

### 3.2 Relation between aESI and gait speed changes.

We first performed an exploratory analysis of the relation between ESI-derived features and the changes in gait speed. Correlation maps between aESI and DGS of all subjects were computed for each method (LORETA, ENETL, and NN-SLASSO). Statistically significant correlations were assessed by a permutation test (50000 permutations) that corrected for multiple comparisons. This analysis was performed at the two different spatial levels, i.e., using voxel-wise and ROI-wise aESI.

Figure 2, panel A, illustrates the voxel-wise correlation map for the three aESI estimation methods (the scale corresponds to the correlation value, restricted to those values with corrected p-value higher than 0.05). LORETA method did not show significant correlations, while the ENET-L and NN-SLASSO methods showed significant differences in the inferior and middle left temporal, left frontal, and occipital regions. A plausible explanation for this result may be the higher spatial smoothness of LORETA with respect to the other methods, which might make the distribution of correlations very flat.

Figure 2, panel B, shows the distribution of correlation values between ROI-wise aESI and the DGS. In each box plot, the horizontal red lines denote the threshold in both tails and display the percentage of the total number of significant ROIs (p<0.05 corrected for multiple comparisons using a permutation test with 50000 permutations). LORETA led to the greatest number of significant ROIs among the three aESI, contrary to the result of no significant voxels found in the voxel-wise correlation analysis. This could be again explained by the low resolution of LORETA, which could lead to a significant level in the ROI analysis among subjects despite low variability in the voxel level. For ENETL and NN-LASSO methods, the two correlation analyses were in accordance. In the case of ENET L, we found seven ROIs with significant negative correlation, with the parahippocampal, fusiform, and temporal middle right regions presenting the highest absolute values. For NN-SLASSO, we localized seven ROIs with significant positive and negative correlations, in which the inferior temporal bilateral (positive correlations) and superior frontal orbital right (negative correlations) regions showed the highest absolute values. Table S2 of the Supplemental Material summarizes the location of the significant regions for each method.

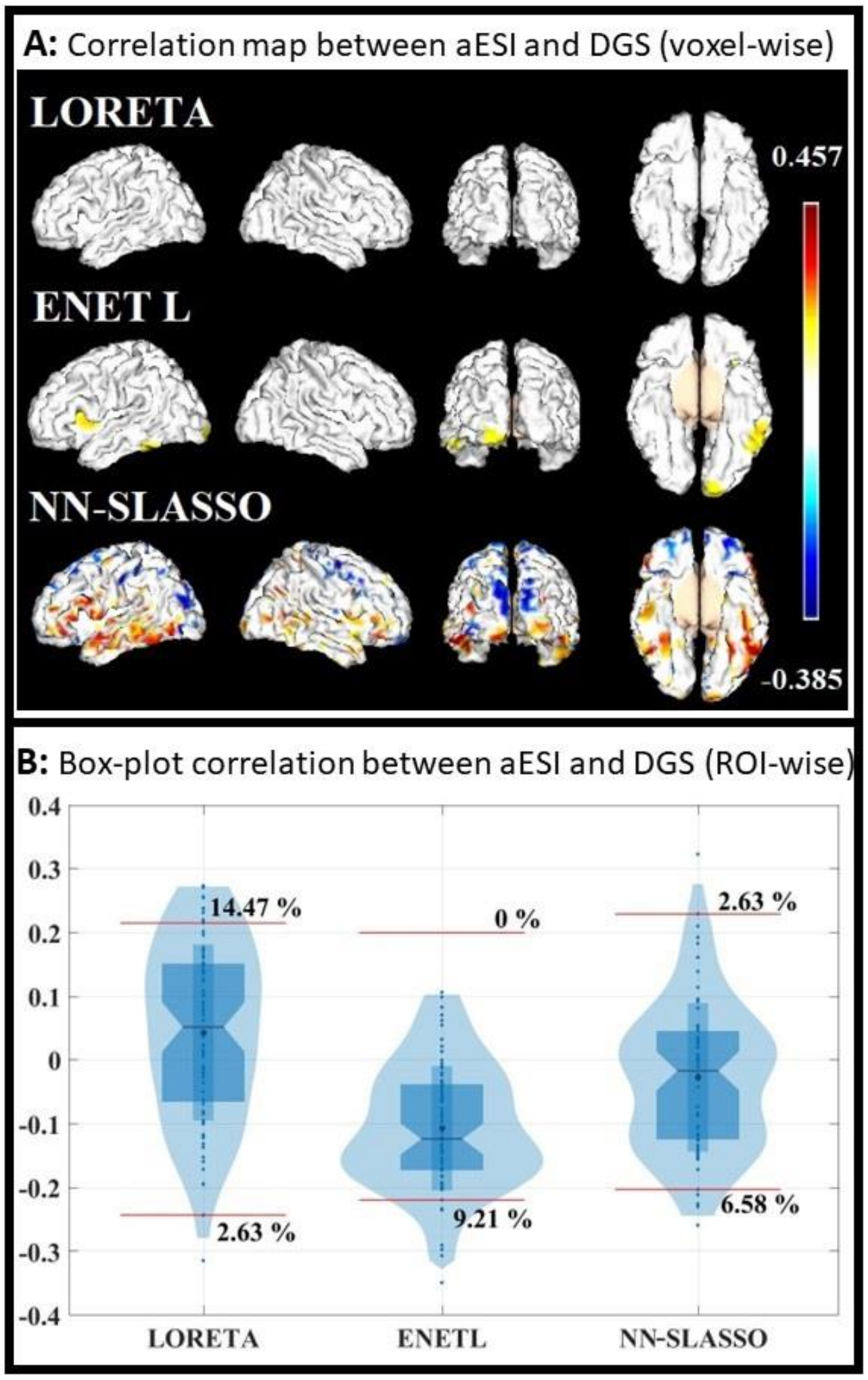

***Fig. 2.*** ***A*) Four views (left, right, back and bottom) images of the significant correlation map between voxel-wise aESI and Delta Gait Speed (DGS). The color bar represents the correlation value, restricted to those values with corrected p-value higher than 0.05. *B)* Box plot of the correlation values between aESI and DGS in ROI-wise analysis. In each tail, we show the significance threshold and the percentage of ROIs in which aESI was significantly correlated with DGS. In both cases, correction for multiple comparisons was performed with 50000 permutations.**

### *3.3 Biomarkers selection from ESI-derived features*

We explored the performance as potential biomarkers of nine adjusted models corresponding to the combination of the three MPLS methods (LORETA/ENET L/NN-SLASSO) and the three ESI-derived features (aESI/cESI/acESI). Figure 3 graphically represents the mean intraclass correlation coefficient (ICC) and its confidence interval (ICC-CI) in each case for the ROI-wise analysis, showing the median ICC value with asterisks. In addition to estimating ICC, a hypothesis test is performed with the null hypothesis ICC = 0. The F value, degrees of freedom, and the corresponding p-value of this test are reported in Table 3, together with the adjusted $R^2$ and its p-value as provided by the SSC.

For all methods, the ICC value was preserved or increased when considering connectivity features compared to activation only. Although all methods and features showed ICC and $R^2$ significantly different from zero, ENET L showed the lowest values for all features and those that changed the least with the inclusion of connectivity. LORETA presented low prediction with mean ICC values below 0.5, although the values increased significantly for cESI and acESI concerning aESI ($p<<0.01$, Table 3). The acESI feature estimated by NN-SLASSO presented the best performance of all cases: the highest mean and smallest variance, with a significant increase when using cESI and acESI with respect to aESI ($p<<0.01$). The acESI from NN-SLASSO may be considered a moderate/suitable predictor of the changes in GS, as reflected by the moderate predictive power $0.50<ICC<0.75$ (Koo & Li, 2016; Liljequist et al., 2019). The adjusted $R^2$ showed the same behavior, although LORETA and ENET L presented similar values and significance, while NN-SLASSO performed better. Table S3 of the Supplemental Material illustrates additional details. These results suggest that the patterns of activation and connectivity obtained from the NN-SLASSO model could be considered possible biomarkers associated with the change in gait speed.

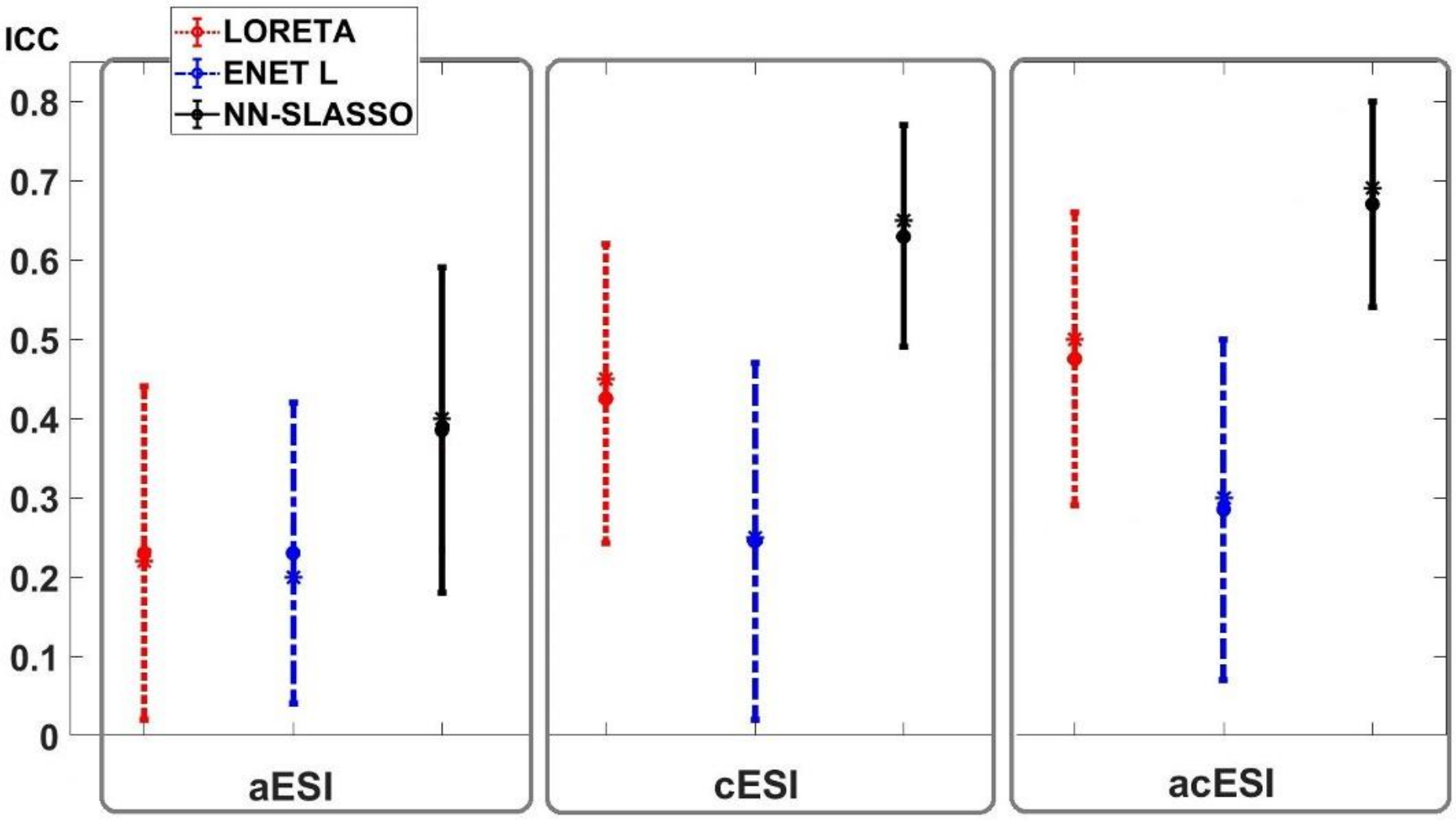

**Fig. 3. The intraclass correlation coefficient (ICC) and its confidence interval (ICC-CI) for the ROI-wise analysis from each feature (aESI, cESI, acESI) and ESI method (LORETA in dotted red line, ENET L in dashed blue line and NN-SLASSO in black solid line) to predict the variation of the gait speed. The median ICC values are represented with asterisks and the mean values with a circle.**

**Table 3.** Performance of models as biomarkers for each feature and ESI model.

| Feature | Method | ICC mean | p-value | $R^2$ | p-value |
|---|---|---|---|---|---|
| aESI | ***LORETA*** | 0.22 | 0.0349 | 0.15 | 0.0021 |
| | ***ENET L*** | 0.45 | 0.0490 | 0.15 | 0.0032 |
| | ***NN-SLASSO*** | 0.50 | 0.0004 | 0.29 | 0.0000 |
| cESI | ***LORETA*** | 0.20 | 0.0001 | 0.33 | 0.0000 |
| | ***ENET L*** | 0.26 | 0.0184 | 0.26 | 0.0001 |
| | ***NN-SLASSO*** | 0.30 | 0.0000 | 0.59 | 0.0000 |
| acESI | ***LORETA*** | 0.40 | 0.0000 | 0.37 | 0.0000 |
| | ***ENET L*** | 0.65 | 0.0065 | 0.27 | 0.0001 |
| | ***NN-SLASSO*** | 0.69 | **0.0000** | **0.62** | **0.0000** |

Figure 4, panel A, shows the activation pattern of the ROI-wise aESI feature, scaled by the value of regression coefficients for each ROI included in the biomarker model. The main ROIs included the superior frontal orbital right, postcentral right, and inferior-temporal bilateral regions. The stability of the coefficients associated with each ROI is in the range of 41.2% and 58.4% (see Table S4 of the Supplemental Material). Panel B shows 149 possible connectivity biomarkers from the cESI feature. Panel C represents the activation map (top) and the connectivity pattern (bottom) associated with the acESI model. Biomarker patterns associated with activation only (Panel A) were highly consistent to those found with combined activation/connectivity, except in the right postcentral region (in blue).

Regarding connectivity, there was also a large agreement between patterns from cESI (Panel B) and the 152 possible connectivity biomarkers estimated for the acESI model. On one hand, the connections between inferior-temporal regions in both hemispheres and between the left frontal inferior triangular and right fusiform regions, shaped the models with the highest replicability (over 98th percentile of the stability index). On the other hand, the connections between the left precuneus and cingulate middle and right posterior regions, and between the left insula and left cuneus, showed the highest positive coefficient values (over the 98th percentile of the absolute value of regression coefficients) in the model. Also, a negative regression coefficient was obtained for a connection between the right supplementary motor area and the right hippocampus.

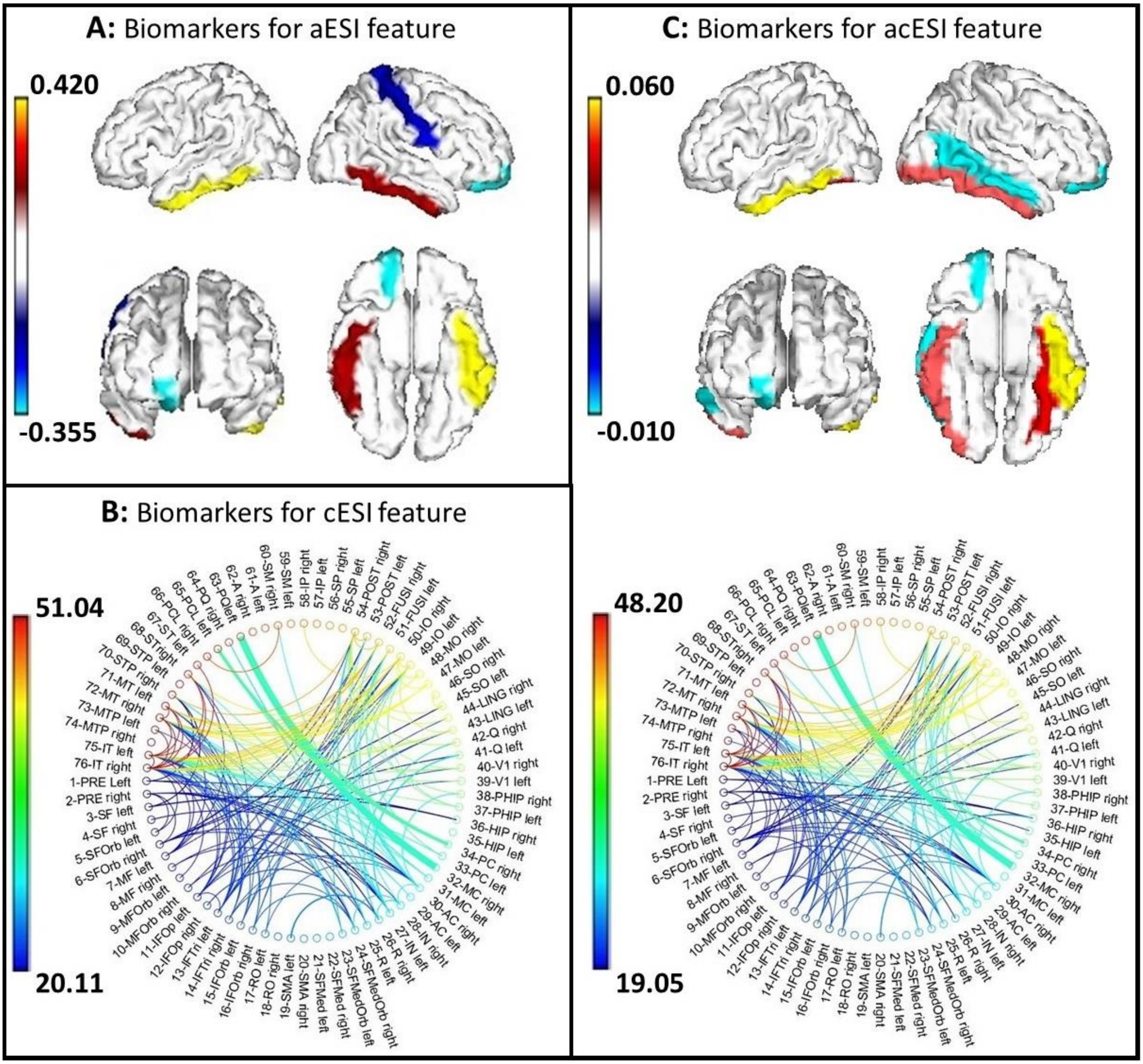

**Fig. 4.** **Candidate activation and connectivity biomarkers for gait speed change. *A)* Four-view (left, right, front, and bottom) activation Electrophysiological Source Imaging (aESI) in which four ROIs are considered biomarkers, with the color bar represents the value of estimated coefficients (right front-orbital superior in cyan, right postcentral in blue, left and right inferior temporal in yellow and red; respectively). *B)* Graph representation of the connectivity Electrophysiological Source Imaging (cESI) in which the width of the connection represents the value of the regression coefficients of the estimated model. The color bar represents the replicability of each variable. *C)* Top panel shows the four views for the five ROIS considered biomarkers from the diagonal of the cross-spectra matrix in the acESI feature (right fronto-orbital superior, left and right inferior temporal, left fusiform and right middle temporal). The color bar represents the value of estimated coefficients. Bottom panel shows a graph representation of the off-diagonal elements of the cross-spectra matrix in the acESI feature. The width of the connection represents the value of the regression coefficients of the estimated model. The color bar represents the replicability of each variable. All variables selected in each model are reported in Tables S4 - S6 of the Supplemental Material.**

## 4. DISCUSSION

Studies on resting-state EEG activation and connectivity patterns have provided essential information about the brain's functional organization and interregional communication. However, selecting early biomarkers of cognitive impairment using joint patterns of activation and connectivity of EEG sources has yet to be well studied and faces validity and applicability challenges. Factors such as the experimental design, the measure chosen as the dependent variables, the spatial resolution of the set of nodes (e.g., region of interest (ROI) vs. single vertices (voxel)) and the methods of analysis can influence the results and limit its validity and potential extension. The variability introduced by EEG recording systems and protocols also affects the replicability of the results. In addition, among the challenges of determining biomarkers are the high dimensionality of ESI features and the need to select a robust prediction method.

Some of these challenges may be addressed with known approaches, but others deserve future thorough studies and creative strategies. For instance, the high variability of evaluation measures can be handled by considering experimental designs from longitudinal studies, which allows the estimation of the basal state inter-individually (Smailovic & Jelic, 2019). Due to the variability introduced by recording protocols, the low replicability can be corrected by applying novel harmonization methods to the collected data (Jovicich et al., 2019; Li et al., 2022). Due to the variable selection method, the variability can be controlled by using performance measures based on the stability of the estimated biomarkers.

The principal outcomes of this paper are to provide a methodology that is easily replicable, to contribute to a better understanding of the role of activation and connectivity ESI patterns as related to gait slowing, and to provide features that can become potential biomarkers of cognitive decline. We applied the proposed methodology to a real data sample from a longitudinal study carried out over nine years in elder subjects (García-Agustin et al., 2020).

### Why using activity and connectivity ESI as features?

In the search for potential biomarkers of GS changes, we explored using the activity and/or connectivity ESI as the independent variable (i.e., as the features). ESI was estimated from three MPLS models with different penalty combinations to evaluate the influence on the predictive power of the estimated activation's degree of smoothness/sparseness (Vega-Hernández et al., 2019). The features and models with the highest predictability and/or stability were selected using the Stable Sparse Classifiers (SSC) procedure (Bosch-Bayard et al., 2018).

In particular, the average amplitude of the individual EEG frequency spectrum in the Theta band was chosen for source estimation using ESI. Previous studies have already considered the Theta band for identifying biomarkers in the study of cognitive decline (Musaeus et al., 2018; Babiloni et al., 2016; Hata et al., 2016; Cozac et al., 2016). . However, most of the studies searching for biomarkers in source activity/connectivity have only used very smooth ESI, such as those obtained with the LORETA family (LORETA, sLORETA, eLORETA) (Pascual-Marqui et al., 1994; Pascual-Marqui, 2002; Pascual-Marqui, 2007). Few reports of studies use sparse ESI or other methods based on the MPLS methodology. In one of these works, Prichep and colleagues used EEG inverse solutions in the frequency domain (VARETA) for the early detection of cognitive decline (Prichep, 2007). Prichep (2007) found similar results to our study by showing that the EEG Theta band predicts future cognitive decline or conversion to dementia with high accuracy. However, they did not include connectivity analysis as a possible biomarker. In a previous preliminary study, we showed that using sparse and sparse/smooth/nonnegativity constraints in combination models for aESI estimation provided more flexibility and adaptability to find complex multisource patterns like those appearing in real data in normal and pathological aging (Vega-Hernández et al., 2022).

In this work, we used the AMNR algorithm that allowed a fast estimation of the different models (Vega-Hernández et al., 2019). We tested the Ridge L solution as a version of LORETA, the ENET L model that combines penalties inducing smoothness and sparsity simultaneously, and the NN-SLASSO, which also combines smoothness and sparsity but adds a nonnegative constraint to the activation. Although the negative values of the activation ESI obtained in a mesh of the cortex can be related to equivalent current dipoles pointing in the opposite direction, the nonnegativity constraint here reflects our prior belief that the most relevant sources correspond to those current dipoles that are similarly oriented (perpendicular to the cortex) with an origin in masses of pyramidal neurons. ESI obtained from these models have shown different abilities to estimate EEG sources with distinct degrees of sparsity. However, the high inter-individual variability of the solution and the lack of a ground truth prevents finding a final answer about the best model to estimate ESI when applied to real EEG data (Vega-Hernández et al., 2022).

We have then tested and compared the predictive ability of activity and connectivity ESI obtained by these three different models, assuming that even if they do not provide the exact true sources in all cases, they do convey replicable patterns that can be related to changes in measures of the clinical decline. For this purpose, we chose the SSC procedure, which performs variable selection with a sparse regression method to handle high-dimensional features (Bosch-Bayard et al., 2018). As an important practical evaluation, we tested the use of voxel-based and ROI-based analyses, since the latter is necessary for decreasing the dimensionality and increasing the robustness of the spatial patterns used as features. Moreover, the SSC procedure combines the regression with resampling techniques that allow the use of non-parametric statistical evaluation of predictability and the stability of the selected variables. The interpretation of penalized regression is relatively straightforward and well-studied, thus mostly accessible (Greenwood et al., 2020). In this context, the methodology presented has broader applicability and can be easily adapted to include other possible predictors and dependent measures of cognitive decline.

### ESI-derived potential biomarkers of gait speed changes

With the proposed methodology, we found that the patterns selected as biomarkers of changes in GS are different according to the ESI model used. For instance, we found that the superior right frontal-orbital region appeared in biomarkers from ENET L and NN-SLASSO, and the right precentral and postcentral regions coincided in patterns obtained with LORETA and NN-SLASSO. The NN-SLASSO model showed the best performance in terms of Inter-Class Correlation (ICC) for all ESI features compared to the other methods. It could be considered a combination of Ridge L (LORETA) and LASSO with the additional sign constraint. We have previously observed that this method is usually sparser than ENET L and provides better localization of deep sources (Vega-Hernández et al., 2019). It is possible that the higher sparsity helped improve the model's predictability since a lower spatial resolution usually masks multiple sources that would provide a more informative spatial pattern. However, highly sparse solutions can also show higher variability in the amplitude and location of the primary sources, which could lead to a higher risk of overfitting and, therefore, lower generalization. In this sense, it is very interesting that models obtained from NN-SLASSO also showed higher replicability values than the other two models. This means that the patterns of sources might be stable enough to provide robust predictability.

The connectivity ESI used as a feature to predict changes in GS was found to be more informative than the activation-only ESI patterns at the ROI level. In all ROI-based models, using features including connectivity (cESI and acESI) enhanced the predictability as measured by the ICC. The best results were obtained from acESI, i.e., when both activation and connectivity information were used as predictors, in this case, by using a vector formed by the variances and absolute values of covariances among all pairs of ROIs. This result supports those previous works claiming a leading role in the pattern of functional connections to explaining changes in cognitive functions, specifically in cognitive decline related to normal and pathological aging. In this work, the main patterns of connections, as potential biomarkers, included the inferior temporal regions in both hemispheres, left inferior frontal triangular and right fusiform areas (with the highest replicability), as well as the connections between the left precuneus and right middle and posterior cingulate regions, and between the left insula and the left cuneus (with the highest positive regression coefficients).

To our knowledge, there are no studies exploring the relation between the sources of resting-state EEG and long gait speed changes. However, some works report the neural correlates of mobility and gait-related measures. Wilson et al., (2019) showed that imaging markers such as white matter integrity and cortical atrophy in frontal areas and basal ganglia showed higher correlation to gait speed than other anatomical and functional markers. However, they did not report results using EEG or EEG sources. Changes in EEG activity at the sensor level have been mainly explored during walking tasks, reporting increases in time-frequency activity for different bands (Huang et al., 2022) which have even been proposed as features to classify and predict walking speed (Rahrooh, 2019).

Several researchers have looked at these neural correlates related to aging. Sanctis et al., (2023) studied the dipole sources obtained from an independent components (IC) analysis of EEG signals recorded during walking on a treadmill in conditions of gait adjustment. Although Sanctis et al., (2023) looked at changes in predefined regions (rather than searching across the whole brain), they found that individuals at higher risk for cognitive impairment showed amplified Theta sources on the right frontal medial and central gyrus. A previous work using sources of IC to study changes according to gait speed also found a relation between activation in prefrontal, posterior parietal, and sensorimotor networks and specific bands (Bulea et al., 2015). These authors concluded that compensatory mechanisms explain the changes in cortical activation in elders with slow gait speed (related to cognitive impairment), which has also been supported by recent studies using NIRS (Greenfield et al., 2023). Despite the methodological differences with the studies mentioned above, our results support the hypothesis that Theta activation localized in the frontal-medial cortex may reflect higher-order compensatory responses to impairments in basic sensorimotor processes.

Finally, we also found that the levels of predictiveness achieved with our models according to the ICC values is similar to those reported in previous studies. For instance, Babiloni et al., (2011) showed moderate Alzheimer's disease predictivity based on source functional connectivity using the LORETA method. However, not all studies performed the same statistical assessment. Therefore, although normalized, the exact values of ICC should not be used as a comparable indicator of the predictive capacity of the features studied.

## Limitations of the study and future work

Our work added further evidence on the usefulness of combining the activation and connectivity derived from ESI as features to predict measures related to cognitive decline. This was especially relevant when using the NN-SLASSO model for obtaining ESI and the SSC methodology to extract robust potential biomarkers. The prediction results and the regions included in the biomarkers are strongly influenced by the ESI model and the method of selection of variables used, as mentioned above. This fact affects the possibility of making a fair comparison and obtaining high replicability with respect to previous results and among them. Future studies should address the limitations in practice to find comparable biomarkers since selecting and comparing them requires careful consideration of various factors to ensure their robustness and generalizability. Also, further research is required to understand a more direct relationship between activation and connectivity EEG as biomarkers for cognitive decline.

Although we used a validated approach to the anatomical ROI definition for specific brain areas, showing excellent longitudinal stability for multiple cortical and subcortical regions (Armstrong et al., 2019), this may also limit comparisons with other papers that use functional or any other type of ROI definitions. This is a factor that will also influence the specific results. Finally, another limitation of our study is the moderate sample size. Future studies with larger cohorts and better sample stratification could also be performed to explore the potential diagnostic role of these features for the early detection of cognitive impairment risk.

## 5. Conclusions

In this study, we selected those stable source activity/connectivity patterns of resting-state EEG theta rhythm as early biomarkers of an indirect measure (gait speed (GS)) of functional decline in aging individuals. For this purpose, we introduced a general methodology for selecting stable biomarkers from activation and connectivity ESI, which allowed the evaluation of ESI models with different constraints. We found that the NN-SLASSO model, that combine smoothness and sparsity with non-negative constraints, outperformed traditional methods, such as the LORETA solution, in predicting longitudinal changes in GS. The features obtained from combining activation and connectivity ESI were the most stable and provided the best prediction of GS changes. Potential biomarkers from activation/connectivity patterns of the EEG theta band involved orbitofrontal and temporal cortical regions. The proposed methodology contributes to the understanding of how activity/connectivity ESI reflects the brain mechanisms underlying the changes in gait speed and provides potential biomarkers of GS decline. This preliminary work opens novel paths to predict the progression of healthy and pathological aging and might allow for ESI-based evaluation of rehabilitation programs.

## Funding

This work was supported by the VLIR-UOS project "A Cuban National School of Neurotechnology for Cognitive Aging" and the National Fund for Science and Innovation of Cuba. CU2017TEA436A103. Partial funding was also provided by the Cuban National Program for Neuroscience and Neurotechnology, Project "Development of disease progression models for brain dysfunctions" PN305LH013-015. This was also supported in part by the Chengdu MOST grant of 2022 under funding No. GH02-00042-HZ and the CNS program of the University of Electronic Sciences and Technology of China (UESTC) under funding No. Y0301902610100201.

## Declaration of Competing Interest

None of the authors have potential conflicts of interest to be disclosed.

## Appendix A. Supplementary material

The online supplementary material contains complementary details on the real data results.

## SUPPLEMENTAL MATERIAL

# Sparse electrophysiological source imaging predicts aging-related gait speed slowing.

**Vega-Hernández, Mayrim[1,2], Galán-García, Lídice[2], Pérez-Hidalgo-Gato, Jhoanna[2], Ontivero-Ortega, Marlis[2], García-Agustin, Daysi[3], García-Reyes, Ronaldo[2], Bosch-Bayard, Jorge[4], Marinazzo, Daniele[5], Martínez-Montes, Eduardo[2,*] and Valdés-Sosa, Pedro A.[1,2,*]**

[1]The Clinical Hospital of Chengdu Brain Science Institute, MOE Key Lab for Neuroinformation, University of Electronic Science and Technology of China, Chengdu, China.

[2]Cuban Center for Neurosciences, Havana, Cuba.

[3]Cuban Centre for Longevity, Ageing and Health Studies, Cuba

[4]Montreal Neurological Institute, Canada.;

[5]Faculty of Psychology and Educational Sciences, Department of Data Analysis, Ghent University.

* Corresponding author at:

Cuban Center for Neurosciences, Human Brain Mapping Division, Havana, Cuba.

The Clinical Hospital of Chengdu Brain Science Institute, MOE Key Lab for Neuroinformation, University of Electronic Science and Technology of China, Chengdu, China.

E-mail address: eduardo@cneuro.edu.cu; pedro.valdes@neuroinformatics-collaboratory.org.

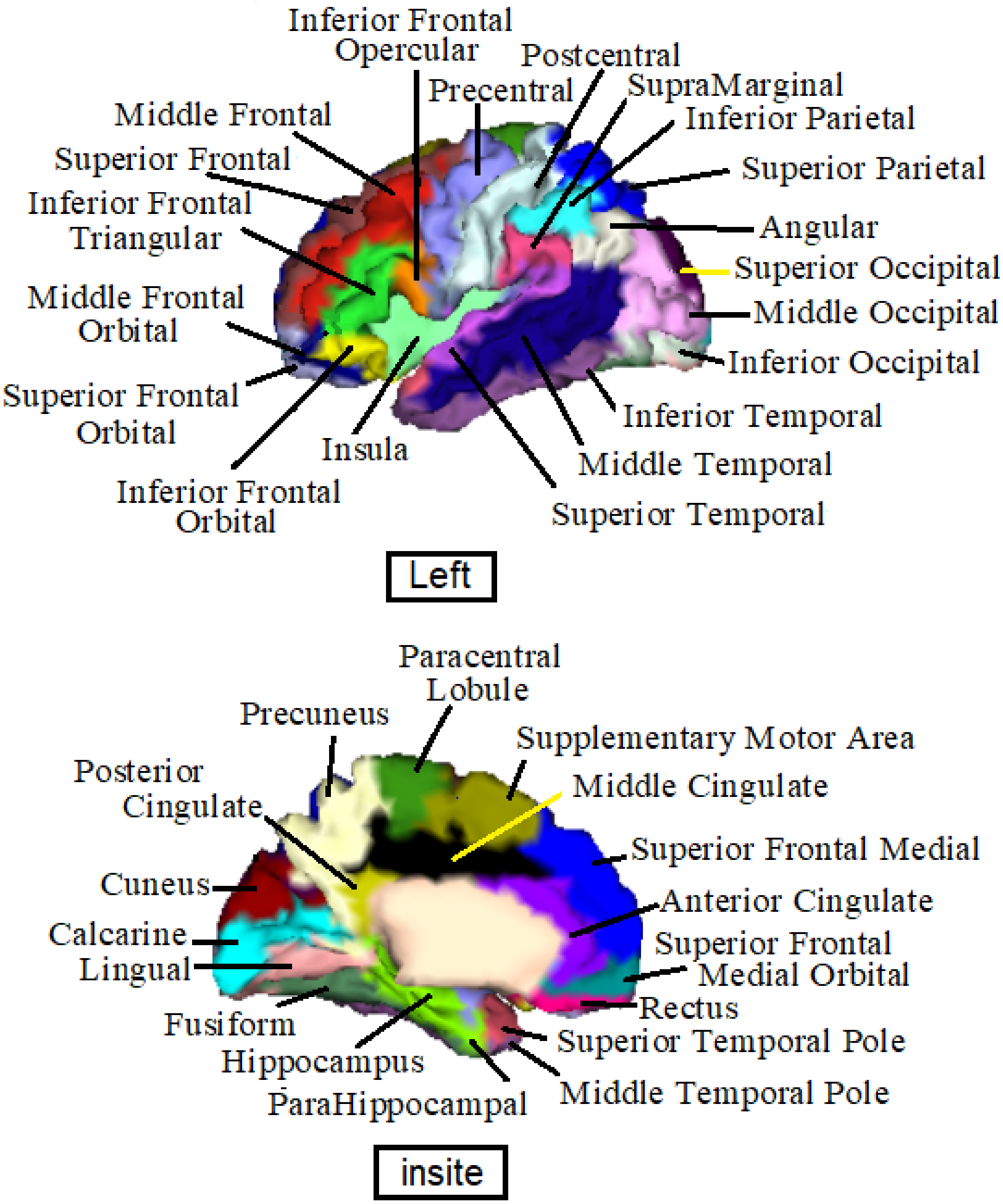

**Fig. S1**: Set of 76 nodes (including left and right hemispheres, as showing in Table S1) selected from anatomical compartments included in the segmentation of the Automated Anatomical Labeling (AAL) average brain of the Montreal Neurological Institute (Tzourio-Mazoyer et al., 2002)

**Table S1.** Index of the 76 cortical labels extracted from the segmentation of the Automated Anatomical Labeling (AAL) average brain of the Montreal Neurological Institute.

| **Name** | **Label** | **Hemisphere** | | **Name** | **Label** | **Hemisphere** | |
|---|---|---|---|---|---|---|---|
| | | **Left** | **Right** | | | **Left** | **Right** |
| Precentral | PRE | 1 | 2 | Calcarine | V1 | 39 | 40 |
| Superior Frontal | SF | 3 | 4 | Cuneus | Q | 41 | 42 |
| Superior Frontal Orbital | SFOrb | 5 | 6 | Lingual | LING | 43 | 44 |
| Middle Frontal | MF | 7 | 8 | Superior Occipital | SO | 45 | 46 |
| Middle Frontal Orbital | MFOrb | 9 | 10 | Middle Occipital | MO | 47 | 48 |
| Inferior Frontal Opercular | IFOp | 11 | 12 | Inferior Occipital | IO | 49 | 50 |
| Inferior Frontal Triangular | IFTri | 13 | 14 | Fusiform | FUSI | 51 | 52 |
| Inferior Frontal Orbital | IFOrb | 15 | 16 | Postcentral | POST | 53 | 54 |
| Rolandic Operculum | RO | 17 | 18 | Superior Parietal | SP | 55 | 56 |
| Supplementary Motor Area | SMA | 19 | 20 | Inferior Parietal | IP | 57 | 58 |
| Superior Frontal Medial | SFMed | 21 | 22 | SupraMarginal | SM | 59 | 60 |
| Superior Frontal Medial Orbital | SFMedOrb | 23 | 24 | Angular | A | 61 | 62 |
| Rectus | R | 25 | 26 | Precuneus | PQ | 63 | 64 |
| Insula | IN | 27 | 28 | Paracentral Lobule | PCL | 65 | 66 |
| Anterior Cingulate | AC | 29 | 30 | Superior Temporal | ST | 67 | 68 |
| Middle Cingulate | MC | 31 | 32 | Superior Temporal Pole | STP | 69 | 70 |
| Posterior Cingulate | PC | 33 | 34 | Middle Temporal | MT | 71 | 72 |
| Hippocampus | HIP | 35 | 36 | Middle Temporal Pole | MTP | 73 | 74 |
| ParaHippocampal | PHIP | 37 | 38 | Inferior Temporal | IT | 75 | 76 |

**Table S2.** ROIs with significative correlation with GS.

| | # | **Label** | ROI Name | Correlation |
|---|---|---|---|---|
| LORETA | 44 | LING | Lingual Right | -0.3147 |
| | 57 | IP | Inferior Parietal left | 0.2736 |
| | 59 | SM | SupraMarginal left | 0.2708 |
| | 53 | POST | Postcentral left | 0.2549 |
| | 55 | SP | Superior Parietal left | 0.2562 |
| | 50 | IO | Inferior Occipital Right | -0.2434 |
| | 18 | RO | Rolandic Operculum Right | 0.2383 |
| | 61 | A | Angular left | 0.2374 |
| | 54 | POST | Postcentral Right | 0.2314 |
| | 2 | PRE | Precentral Right | 0.2200 |
| | 63 | PQ | Precuneus left | 0.2157 |
| | 65 | PCL | Paracentral Lobule left | 0.2153 |
| | 67 | ST | Superior Temporal left | 0.2151 |
| | | | | |
| ENET L | 38 | PHIP | ParaHippocampal Right | -0.3492 |
| | 72 | MT | Middle Temporal Right | -0.3067 |
| | 52 | FUSI | Fusiform Right | -0.2971 |
| | 36 | HIP | Hippocampus Right | -0.2906 |
| | 74 | MTP | Middle Temporal Pole Right | -0.2353 |
| | 26 | R | Rectus Right | -0.2330 |
| | 6 | SFOrb | Superior Frontal Orbital Right | -0.2194 |
| | | | | |
| NN-SLASSO | 75 | | Inferior Temporal left | 0.3230 |
| | 6 | SFOrb | Superior Frontal Orbital Right | -0.2590 |
| | 54 | POST | Postcentral Right | -0.2300 |
| | 76 | IT | Inferior Temporal Right | 0.2293 |
| | 10 | MFOrb | Middle Frontal Orbital Right | -0.2253 |
| | 30 | AC | Anterior Cingulate Right | -0.2112 |
| | 2 | PRE | Precentral Right | 0.2033 |

**Table S3.** Performance of models for each feature (aESI, cESI, acESI) and each method (LORETA, ENET L, NN-SLASSO) tested

| Method | Feature | ICC | ICC-IC | p-value | F | R | p |
|---|---|---|---|---|---|---|---|
| **LORETA** | **aESI** | 0.2220 | [0.0184 0.4384] | 0.0349 | 1.5706 | 0.15 | 0.0021 |
| | **cESI** | 0.4508 | [0.2371 0.6232] | 5.9468e-05 | 2.6418 | 0.33 | 0.0000 |
| | **acESI** | 0.4967 | [0.2921 0.6581] | 8.6236e-06 | 2.9738 | 0.37 | 0.0000 |
| **ENET L** | **aESI** | 0.2029 | [0.0383 0.4222] | 0.0490 | 1.5092 | 0.15 | 0.0032 |
| | **cESI** | 0.2629 | [0.0252 0.4729] | 0.0154 | 1.7134 | 0.26 | 0.0000 |
| | **acESI** | 0.3009 | [0.0663 0.5043] | 0.0065 | 1.8608 | 0.27 | 0.0000 |
| **NN-SLASSO** | **aESI** | 0.4016 | [0.1795 0.5850] | 0.0004 | 2.3423 | 0.29 | 0.0000 |
| | **cESI** | 0.6545 | [0.4923 0.7730] | 7.6453e-10 | 4.7884 | 0.59 | 0.0000 |
| | **acESI** | 0.6878 | [0.5369 0.7963] | 4.9805e-11 | 5.4068 | 0.62 | 0.0000 |

**Table S4.** Selected models based on aESI feature estimated by NN-SLASSO method.

| aESI | | | |
|---|---|---|---|
| **#** | **ROI Name** | **Replicability** | **Beta in Best Model** |
| 75 | Inferior Temporal Left | 58,38 | 0,42 |
| 54 | Postcentral Right | 42,50 | -0,20 |
| 76 | Inferior Temporal Right | 41,21 | 0,23 |
| 6 | Superior Frontal Orbital Right | 41,18 | -0,36 |

**Table S5.** Selected models based on cESI feature estimated by NN-SLASSO method.

| cESI | | | |
|---|---|---|---|
| **Name ROI 1** | **Name ROI 2** | **Replicability** | **Beta in Best Model** |
| Middle Cingulate Right | Precuneus Left | 20,11 | 13,95 |
| Posterior Cingulate Right | Paracentral Lobule Left | 20,59 | 6,01 |
| Posterior Cingulate Right | Precuneus Left | 26,28 | 4,67 |
| Insula Left | Cuneus Left | 22,31 | 1,57 |
| Middle Cingulate Left | Paracentral Lobule Left | 21,01 | 1,11 |
| Middle Cingulate Right | Inferior Temporal Left | 22,25 | 0,86 |
| Posterior Cingulate Left | Inferior Temporal Left | 24,37 | 0,79 |
| Posterior Cingulate Left | Fusiform Left | 24,07 | 0,59 |
| Middle Cingulate Left | Inferior Temporal Left | 20,93 | 0,51 |
| Rolandic Operculum Left | Middle Temporal Right | 22,35 | 0,30 |
| Inferior Frontal Orbital Right | Inferior Temporal Left | 27,03 | 0,27 |
| Inferior Frontal Opercular Left | Insula Right | 20,46 | 0,27 |
| Inferior Frontal Triangular Left | Fusiform Right | 47,73 | 0,27 |
| Anterior Cingulate Left | Inferior Temporal Left | 23,01 | 0,24 |
| Inferior Frontal Orbital Left | Lingual Left | 22,10 | 0,22 |
| Insula Right | Fusiform Left | 30,53 | 0,22 |
| Inferior Frontal Opercular Left | Middle Temporal Left | 36,81 | 0,21 |
| Rectus Left | Middle Temporal Right | 22,74 | 0,20 |
| Inferior Temporal Left | Inferior Temporal Right | 49,57 | 0,20 |
| Inferior Frontal Triangular Right | Middle Temporal Pole Left | 30,14 | 0,19 |
| Inferior Frontal Orbital Left | Inferior Temporal Right | 46,31 | 0,18 |
| Middle Temporal Right | Inferior Temporal Left | 49,58 | 0,17 |
| Middle Frontal Orbital Left | Inferior Temporal Right | 25,49 | 0,17 |
| Middle Frontal Orbital Left | Inferior Temporal Left | 35,93 | 0,16 |
| Middle Frontal Left | Superior Temporal Right | 29,04 | 0,16 |
| Middle Frontal Left | Middle Temporal Right | 38,32 | 0,15 |
| Inferior Occipital Left | Superior Temporal Right | 23,38 | 0,15 |
| Middle Temporal Pole Left | Inferior Temporal Right | 31,58 | 0,15 |
| Inferior Frontal Orbital Right | Inferior Temporal Right | 23,60 | 0,15 |
| Superior Frontal Orbital Left | Inferior Temporal Right | 22,25 | 0,14 |
| Anterior Cingulate Left | Middle Temporal Right | 21,08 | 0,14 |
| Inferior Frontal Opercular Left | Inferior Temporal Left | 33,43 | 0,14 |
| Superior Temporal Right | Inferior Temporal Right | 24,43 | 0,13 |
| Inferior Frontal Triangular Right | Inferior Temporal Left | 38,62 | 0,13 |
| Lingual Left | Inferior Temporal Right | 36,84 | 0,13 |
| ParaHippocampal Left | Inferior Occipital Left | 22,10 | 0,13 |
| Inferior Occipital Left | Inferior Temporal Right | 43,91 | 0,12 |
| Middle Frontal Orbital Left | Inferior Temporal Right | 33,53 | 0,12 |
| Inferior Frontal Triangular Left | Inferior Temporal Right | 43,79 | 0,11 |
| Inferior Occipital Right | Middle Temporal Left | 27,79 | 0,11 |
| Inferior Occipital Right | Inferior Parietal Right | 22,29 | 0,11 |
| Middle Occipital Right | Inferior Occipital Right | 23,35 | 0,11 |
| Inferior Frontal Triangular Left | Middle Temporal Right | 46,15 | 0,11 |
| Inferior Frontal Triangular Left | Middle Temporal Left | 32,01 | 0,11 |
| Superior Temporal Right | Inferior Temporal Left | 24,93 | 0,11 |
| Inferior Frontal Opercular Left | Middle Temporal Right | 30,14 | 0,11 |
| LingualLeft | Middle Temporal Right | 38,20 | 0,11 |
| Middle Frontal Left | Inferior Temporal Left | 32,49 | 0,10 |
| Inferior Frontal Orbital Left | Middle Temporal Right | 30,54 | 0,10 |
| Inferior Frontal Orbital Right | Inferior Occipital Left | 21,67 | 0,10 |
| Inferior Frontal Triangular Left | Lingual Right | 28,32 | 0,10 |
| Inferior Frontal Triangular Left | Superior Temporal Right | 25,80 | 0,10 |
| Lingual Left | Inferior Temporal Left | 28,53 | 0,10 |
| Inferior Frontal Orbital Right | Fusiform Left | 20,56 | 0,10 |
| Calcarine Right | Inferior Temporal Left | 28,33 | 0,10 |
| Middle Temporal Left | Inferior Temporal Right | 32,26 | 0,10 |
| Inferior Frontal Triangular Left | Inferior Occipital Right | 51,04 | 0,10 |
| Middle Occipital Left | Inferior Occipital Right | 31,21 | 0,09 |
| Middle Occipital Left | Inferior Temporal Left | 35,94 | 0,09 |
| Inferior Frontal Triangular Left | Inferior Temporal Left | 34,69 | 0,09 |
| Fusiform Right | Middle Temporal Left | 25,95 | 0,09 |
| Middle Frontal Orbital Left | Middle Temporal Right | 24,50 | 0,09 |
| Superior Temporal Left | Inferior Temporal Right | 30,19 | 0,08 |
| Inferior Frontal Orbital Left | Inferior Occipital Right | 23,91 | 0,08 |
| Inferior Frontal Opercular Left | Inferior Temporal Right | 28,65 | 0,08 |
| Middle Frontal Orbital Left | Inferior Occipital Left | 28,99 | 0,08 |
| Middle Frontal Left | Inferior Temporal Right | 26,27 | 0,07 |
| Inferior Frontal Tri Left | Lingual Left | 30,81 | 0,07 |
| Middle Occipital Right | Middle Temporal Left | 23,93 | 0,07 |
| Calcarine Left | Inferior Temporal Right | 24,86 | 0,07 |
| Inferior Frontal Opercular Left | Calcarine Right | 23,14 | 0,07 |
| Middle Occipital Left | Inferior Temporal Right | 26,68 | 0,06 |
| Inferior Occipital Left | Middle Temporal Right | 41,41 | 0,06 |
| Fusiform Right | Inferior Temporal Right | 26,55 | 0,06 |
| Middle Temporal Left | Inferior Temporal Left | 31,88 | 0,06 |
| 'Fusiform Left | Inferior Temporal Left | 30,84 | 0,06 |
| Inferior Occipital Left | Inferior Temporal Left | 48,71 | 0,05 |
| Middle Occipital Left | Middle Temporal Left | 23,10 | 0,05 |
| Middle Temporal Right | Inferior Temporal Right | 31,55 | 0,05 |
| Middle Occipital Left | Middle Temporal Right | 23,65 | 0,05 |
| Inferior Frontal Triangular Left | Calcarine Left | 20,29 | 0,05 |
| Fusiform Right | Inferior Temporal L' | 28,69 | 0,05 |
| Inferior Frontal Triangular Right | Inferior Temporal Right | 25,94 | 0,05 |
| Inferior Frontal Opercular Left | Inferior Occipital Right | 31,84 | 0,04 |
| Middle Frontal Left | Inferior Occipital Left | 21,73 | 0,03 |
| ParaHippocampal Left | Fusiform Left | 23,63 | 0,03 |
| Fusiform Right | Middle Temporal Right | 21,10 | 0,03 |
| Inferior Frontal Tri Left | Fusiform Left | 35,08 | 0,03 |
| Inferior Occipital Right | Inferior Temporal Left | 37,08 | 0,03 |
| 'Lingual Left | Inferior Occipital Left | 23,10 | 0,03 |
| Inferior Occipital Left | Inferior Occipital Right | 34,94 | 0,03 |
| Inferior Frontal Opercular Left | Inferior Frontal Triangular Left | 27,75 | 0,03 |
| ParaHippocampal Left | Inferior Temporal Left | 26,86 | 0,02 |
| Inferior Frontal Opercular Left | Fusiform Left | 27,09 | 0,02 |
| Inferior Occipital Right | Inferior Temporal Right | 22,10 | 0,02 |
| Calcarine Right | Middle Temporal Left | 22,09 | 0,01 |
| Inferior Occipital Right | Middle Temporal Right | 21,13 | 0,01 |
| Middle Frontal Orbital Right | Inferior Temporal Left | 20,44 | 0,00 |
| Middle Temporal Left | Middle Temporal Right | 26,96 | -0,04 |
| Cuneus Right | Postcentral Right | 20,69 | -0,04 |
| Postcentral Left | Postcentral Right | 20,39 | -0,06 |
| Calcarine Right | Postcentral Right | 24,32 | -0,07 |
| Precentral Right | Anterior Cingulate Right | 21,79 | -0,09 |
| Middle Frontal Orbital Right | Postcentral Left | 20,33 | -0,11 |
| Inferior Frontal Triangular Right | Middle Frontal Orbital Right | 24,78 | -0,11 |
| Anterior Cingulate Right | Postcentral Left | 24,86 | -0,12 |
| Superior Occipital Left | Postcentral Right | 20,23 | -0,12 |
| Middle Frontal Right | Anterior Cingulate Right | 21,17 | -0,12 |
| Superior Frontal Orbital Right | Middle Frontal Orbital Right | 22,38 | -0,13 |
| Lingual Right | Postcentral Right | 22,57 | -0,13 |
| Superior Frontal Orbital Right | Postcentral Left | 26,22 | -0,13 |
| Fusiform Right | Postcentral Right | 20,56 | -0,15 |
| Postcentral Right | Superior Temporal Pole Right | 20,34 | -0,17 |
| Precentral Right | Superior Frontal Orbital Left | 20,61 | -0,17 |
| Middle Frontal Right | Middle Frontal Orbital Right | 24,93 | -0,17 |
| Anterior Cingulate Right | Inferior Occipital Left | 21,62 | -0,18 |
| Rectus Left | Postcentral Right | 27,42 | -0,22 |
| Superior Occipital Left | Superior Occipital Right | 20,41 | -0,22 |
| Superior Frontal Orbital Right | Postcentral Right | 34,14 | -0,23 |
| Insula Right | Anterior Cingulate Right | 24,36 | -0,23 |
| Inferior Frontal Triangular Right | Anterior Cingulate Right | 28,32 | -0,24 |
| Precentral Right | Superior Frontal Orbital Right | 26,45 | -0,24 |
| Anterior Cingulate Right | SupraMarginal Right | 25,07 | -0,27 |
| Middle Frontal Orbital Right | Insula Right | 22,37 | -0,27 |
| Precentral Right | Rectus Left | 24,36 | -0,28 |
| Anterior Cingulate Right | Superior Temporal Right | 23,77 | -0,29 |
| Middle Frontal Orbital Right | Insula Left | 23,12 | -0,30 |
| Precentral Right | Rectus Right | 21,99 | -0,34 |
| Rolandic Operculum Left | Anterior Cingulate Right | 23,85 | -0,36 |
| ParaHippocampal Right | 'Postcentral Right | 25,27 | -0,37 |
| Rectus Right | Insula Right | 20,66 | -0,37 |
| SupraMarginal Right | Paracentral Lobule Right | 20,16 | -0,37 |
| Precentral Left | Hippocampus Right | 27,17 | -0,40 |
| Precentral Left | Rectus Right | 24,79 | -0,41 |
| Superior Frontal Orbital Right | Insula Left | 25,00 | -0,42 |
| Anterior Cingulate Right | Fusiform Right | 24,85 | -0,44 |
| Inferior Frontal Orbital Right | Middle Frontal Orbital Right | 29,45 | -0,48 |
| Precentral Left | Superior Temporal Pole Right | 21,07 | -0,50 |
| Inferior Frontal Opercular Right | Rectus Left | 29,48 | -0,53 |
| Superior Frontal Orbital Right | Cuneus Left | 26,91 | -0,57 |
| Superior Frontal Right | Anterior Cingulate Right | 24,02 | -0,59 |
| Insula Left | Superior Temporal Pole Right | 23,97 | -0,60 |
| Rectus Right | Cuneus Right | 20,53 | -0,63 |
| Anterior Cingulate Right | Fusiform Left | 32,38 | -0,65 |
| Anterior Cingulate Right | ParaHippocampal Right | 26,45 | -0,67 |
| Insula Left | Precuneus Right | 21,08 | -0,69 |
| Superior Frontal Orbital Right | Superior Occipital Left | 25,15 | -0,69 |
| Precentral Right | Insula Left | 22,06 | -0,69 |
| Superior Frontal Left | Superior Temporal Pole Right | 24,43 | -0,78 |
| Insula Left | Postcentral Right | 23,25 | -0,83 |
| Superior Frontal Left | Hippocampus Right | 27,68 | -1,16 |
| Supplementary Motor Area Left | Rectus Right | 20,23 | -1,71 |

**Table S6.** Selected models based on acESI feature estimated by NN-SLASSO method.

| **acESI** | | | |
|---|---|---|---|
| **Name ROI 1** | **Name ROI 2** | **Replicability** | **Beta in Best Model** |
| Middle Cingulate Right | Precuneus Left | 22,47 | **19,69** |
| Posterior Cingulate Right | Precuneus Left | 24,31 | **6,08** |
| Insula Left | Cuneus Left | 24,43 | **1,42** |
| Posterior Cingulate Left | Paracentral Lobule Left | 22,34 | **1,33** |
| Middle Cingulate Right | Inferior Temporal Left | 24,11 | **1,28** |
| Middle Cingulate Right | Inferior Occipital Right | 22,41 | 0,87 |
| Posterior Cingulate Left | Inferior Temporal Left | 23,23 | 0,82 |
| Rolandic Operculum Left | Middle Temporal Pole Right | 21,39 | 0,80 |
| Posterior Cingulate Left | Fusiform Left | 19,88 | 0,67 |
| Inferior Frontal Orbital Left | Lingual Left | 19,60 | 0,32 |
| Inferior Frontal Orbital Right | Inferior Temporal Left | 23,33 | 0,30 |
| Rolandic Operculum Left | Middle Temporal Right | 22,62 | 0,29 |
| Inferior Frontal Triangular Right | Middle Temporal Pole Left | 22,07 | 0,29 |
| **Inferior Frontal Triangular Left** | **Fusiform Right** | **47,21** | 0,28 |
| Anterior Cingulate Left | Inferior Temporal Left | 19,71 | 0,27 |
| Insula Right | Fusiform Left | 27,65 | 0,26 |
| Inferior Frontal Opercular Left | Middle Temporal Left | 33,62 | 0,25 |
| Inferior Frontal Orbital Left | Inferior Temporal Right | 39,07 | 0,21 |
| Middle Frontal Orbital Left | Inferior Temporal Left | 28,41 | 0,20 |
| Middle Frontal Orbital Left | Inferior Temporal Right | 24,48 | 0,19 |
| Rectus Left | Middle Temporal Right | 22,81 | 0,19 |
| Middle Frontal Left | Middle Temporal Right | 34,92 | 0,18 |
| Inferior Frontal Opercular Left | Inferior Temporal Left | 29,04 | 0,18 |
| **Inferior Temporal Left** | **Inferior Temporal Right** | **48,20** | 0,17 |
| **Middle Temporal Right** | **Inferior Temporal Left** | **40,71** | 0,17 |
| Inferior Occipital Left | Superior Temporal Right | 25,93 | 0,17 |
| Middle Frontal Left | Superior Temporal Right | 27,07 | 0,16 |
| Inferior Frontal Opercular Left | Middle Temporal Right | 25,80 | 0,15 |
| Middle Temporal Pole Left | Inferior Temporal Right | 28,49 | 0,15 |
| Middle Frontal Left | Inferior Temporal Left | 26,09 | 0,14 |
| Superior Frontal Orbital Left | Inferior Temporal Right | 19,94 | 0,14 |
| Inferior Frontal Triangular Right | Inferior Temporal Left | 35,92 | 0,13 |
| Lingual Left | Inferior Temporal Right | 34,08 | 0,13 |
| Superior Temporal Right | Inferior Temporal Right | 27,51 | 0,12 |
| **Inferior Frontal Triangular Left** | **Middle Temporal Right** | **43,02** | 0,12 |
| Lingual Left | Middle Temporal Right | 37,03 | 0,12 |
| **Inferior Frontal Triangular Left** | **Inferior Temporal Right** | **41,85** | 0,12 |
| Lingual Left | Inferior Temporal Left | 24,64 | 0,12 |
| Inferior Frontal Triangular Left | Lingual Right | 21,93 | 0,11 |
| **Inferior Occipital Left** | **Inferior Temporal Right** | **42,69** | 0,11 |
| Inferior Frontal Orbital Left | Middle Temporal Right | 30,73 | 0,11 |
| Inferior Occipital Right | Inferior Parietal Right | 23,21 | 0,11 |
| Inferior Frontal Opercular Left | Inferior Temporal Right | 24,32 | 0,11 |
| Anterior Cingulate Left | Middle Temporal Right | 20,11 | 0,10 |
| Inferior Frontal Triangular Left | Superior Temporal Right | 24,86 | 0,10 |
| Middle OccipitalRight | Inferior Occipital Right | 24,57 | 0,10 |
| Inferior Frontal Triangular Left | Lingual Left | 27,00 | 0,10 |
| Inferior Frontal Triangular Left | Middle Temporal Left | 29,23 | 0,10 |
| Middle Frontal Right | Inferior Temporal Right | 31,01 | 0,10 |
| Middle Frontal Orbital Left | Middle Temporal Right | 21,49 | 0,09 |
| Calcarine Right | Inferior Temporal Left | 26,15 | 0,09 |
| Middle Frontal Orbital Left | Inferior Occipital Left | 22,79 | 0,09 |
| Superior Temporal Left | Inferior Temporal Right | 24,57 | 0,09 |
| Inferior Occipital Right | Middle Temporal Left | 26,72 | 0,09 |
| Inferior Frontal Triangular Left | Fusiform Left | 25,93 | 0,09 |
| Middle Occipital Left | Inferior Temporal Left | 32,84 | 0,09 |
| Inferior Frontal Opercular Left | Calcarine Right | 22,73 | 0,09 |
| Middle Frontal Left | Inferior Temporal Right | 20,62 | 0,08 |
| Middle Occipital Left | Inferior Occipital Right | 31,86 | 0,08 |
| Fusiform Right | Middle Temporal Left | 21,61 | 0,08 |
| **Inferior Frontal Triangular Left** | **Inferior Occipital Right** | **45,78** | 0,08 |
| Middle Temporal Left | Inferior Temporal Right | 32,58 | 0,08 |
| Superior Temporal Right | Inferior Temporal Left | 24,79 | 0,08 |
| Inferior Frontal Opercular Left | Fusiform Left | 23,84 | 0,08 |
| Calcarine Left | Inferior Temporal Right | 24,79 | 0,08 |
| Middle Temporal Left | Inferior Temporal Left | 24,49 | 0,07 |
| **Inferior Occipital Left** | **Middle Temporal Right** | **40,46** | 0,07 |
| Inferior Frontal Orbital Left | Inferior Occipital Right | 23,12 | 0,07 |
| **Inferior Temporal Left** | **Inferior Temporal Left** | **46,91** | 0,06 |
| Fusiform Left | Inferior Temporal Left | 28,07 | 0,06 |
| Inferior Occipital Left | Inferior Temporal Left | 38,26 | 0,06 |
| Calcarine Left | Inferior Temporal Left | 21,97 | 0,05 |
| Middle Occipital Left | Inferior Temporal Right | 23,80 | 0,05 |
| Middle Occipital Right | Middle Temporal Left | 27,81 | 0,05 |
| Calcarine Right | Inferior Occipital Right | 20,61 | 0,05 |
| Fusiform Right | Inferior Temporal Left | 23,67 | 0,05 |
| Fusiform Right | Inferior Temporal Right | 24,22 | 0,05 |
| Middle Occipital Left | Middle Temporal Left | 20,88 | 0,05 |
| Inferior Frontal Triangular Left | Inferior Temporal Left | 32,34 | 0,05 |
| Middle Temporal Right | Inferior Temporal Right | 31,36 | 0,04 |
| Middle Occipital Left | Middle Temporal Right | 19,37 | 0,04 |
| Inferior Frontal Opercular Left | Inferior Occipital Right | 29,38 | 0,04 |
| Inferior Frontal Triangular Right | Inferior Temporal Right | 21,29 | 0,04 |
| Lingual Left | Inferior Occipital Left | 22,19 | 0,03 |
| Inferior Occipital Right | Superior Temporal Right | 20,49 | 0,03 |
| Inferior Occipital Right | Inferior Temporal Left | 35,95 | 0,03 |
| Inferior Occipital Left | Inferior Occipital Right | 35,17 | 0,03 |
| Inferior Occipital Right | Inferior Temporal Right | 20,51 | 0,02 |
| Fusiform Left | Fusiform Left | 24,71 | 0,02 |
| Inferior Temporal Right | Inferior Temporal Right | 26,44 | 0,01 |
| Calcarine Right | Middle Temporal Left | 19,05 | 0,01 |
| Inferior Occipital Right | Inferior Occipital Right | 21,13 | 0,01 |
| Middle Occipital Right | Middle Temporal Right | 20,23 | -0,01 |
| Middle Frontal Right | Middle Frontal Orbital Right | 19,94 | -0,01 |
| Middle Temporal Right | Middle Temporal Right | 23,86 | -0,01 |
| Superior Frontal Orbital Right | Superior Frontal Orbital Right | 19,21 | -0,01 |
| ParaHippocampal Left | Inferior Temporal Left | 22,95 | -0,03 |
| Fusiform Right | Middle Temporal Right | 19,06 | -0,03 |
| Inferior Occipital Right | Middle Temporal Right | 22,92 | -0,03 |
| Superior Frontal Orbital Right | Anterior Cingulate Right | 21,45 | -0,04 |
| Middle Frontal Right | Inferior Frontal Orbital Right | 20,52 | -0,05 |
| Precentral Right | Anterior Cingulate Right | 20,76 | -0,06 |
| Postcentral Left | Postcentral Right | 19,09 | -0,07 |
| Anterior Cingulate Right | Postcentral Left | 24,85 | -0,08 |
| Superior Frontal Orbital Right | Rolandic Operculum Left | 21,70 | -0,08 |
| Middle Temporal Left | Middle Temporal Right | 21,70 | -0,09 |
| Superior Frontal Orbital Right | Middle Frontal Orbital Right | 24,86 | -0,10 |
| Middle Frontal Right | Anterior Cingulate Right | 19,35 | -0,10 |
| Calcarine Right | Postcentral Right | 23,91 | -0,11 |
| Inferior Frontal Triangular Right | Middle Frontal Orbital Right | 26,04 | -0,12 |
| Superior Frontal Orbital Right | Postcentral Left | 23,44 | -0,13 |
| Precentral Right | Superior Frontal Orbital Right | 29,34 | -0,17 |
| Lingual Right | Postcentral Right | 21,39 | -0,18 |
| Middle Frontal Right | Middle Frontal Orbital Right | 24,49 | -0,19 |
| Superior Frontal Orbital Left | Postcentral Left | 21,27 | -0,22 |
| Insula Right | Anterior Cingulate Right | 22,29 | -0,23 |
| Superior Frontal Orbital Right | Postcentral Right | 30,97 | -0,23 |
| Postcentral Right | Superior Temporal Pole Right | 19,61 | -0,25 |
| Rectus Left | Postcentral Right | 24,02 | -0,26 |
| Middle Frontal Orbital Right | Insula Right | 25,00 | -0,26 |
| Anterior Cingulate Right | SupraMarginal Right | 20,49 | -0,28 |
| Inferior Frontal Triangular Right | Anterior Cingulate Right | 26,39 | -0,28 |
| Precentral Right | Middle Temporal Pole Left | 21,30 | -0,29 |
| Rolandic Operculum Left | Anterior Cingulate Right | 23,86 | -0,29 |
| Superior Frontal Left | Hippocampus Left | 20,34 | -0,30 |
| Precentral Right | Rectus Left | 20,89 | -0,30 |
| Rectus Left | Postcentral Left | 19,37 | -0,35 |
| Precentral Right | Rectus Right | 20,61 | -0,37 |
| Superior Frontal Orbital Right | Insula Left | 25,59 | -0,43 |
| SupraMarginal Right | Paracentral Lobule Right | 19,23 | -0,44 |
| Anterior Cingulate Right | Superior Temporal Left | 21,63 | -0,46 |
| ParaHippocampal Right | Postcentral Right | 23,71 | -0,47 |
| Superior Frontal Right | Anterior Cingulate Right | 22,57 | -0,47 |
| Inferior Frontal Orbital Right | Middle Frontal Orbital Right | 31,18 | -0,48 |
| Anterior Cingulate Right | Fusiform Right | 20,61 | -0,50 |
| Precentral Left | Hippocampus Right | 23,01 | -0,51 |
| Middle Cingulate Right | Postcentral Left | 22,16 | -0,57 |
| Precentral Left | Rectus Right | 22,16 | -0,57 |
| Precentral Right | Insula Left | 23,63 | -0,60 |
| Inferior Frontal Opercular Right | Rectus Left | 28,74 | -0,61 |
| Insula Left | Superior Temporal Pole Right | 23,63 | -0,71 |
| Anterior Cingulate Right | Fusiform Left | 30,34 | -0,72 |
| Superior Frontal Orbital Right | Cuneus Left | 22,51 | -0,78 |
| Superior Frontal Orbital Right | Superior Occipital Left | 27,35 | -0,79 |
| Anterior Cingulate Right | ParaHippocampal Right | 22,81 | -0,85 |
| Superior Frontal Left | Superior Temporal Pole Right | 22,67 | -0,91 |
| Insula Left | Postcentral Right | 23,36 | -1,00 |
| Superior Frontal Left | Hippocampus Right | 29,89 | -1,08 |
| Supplementary Motor Area Right | Hippocampus Right | 19,28 | -1,95 |